\documentclass[journal]{IEEEtran}
\ifCLASSINFOpdf
\else
\fi
\usepackage{tabularx}
\usepackage{makecell}
\usepackage{graphicx}
\usepackage{graphics}
\usepackage{subfig}
\usepackage{multirow}

\usepackage{caption}
\usepackage[font=footnotesize]{caption}
\usepackage{amsmath}
\usepackage[switch,pagewise]{lineno} 
\usepackage{lipsum} 
\usepackage{diagbox}

\begin{document}
%
\title{Disengagement Cause-and-Effect Relationships Extraction Using an NLP Pipeline}

%
%
%

\author{{Yangtao Zhang, X. Jessie Yang, Feng Zhou}
\thanks{Y. Zhang is with School of Information, The University of Michigan, Ann Arbor, 500 S State St, Ann Arbor, MI 48109 USA (e-mail: maxzhang@umich.edu).}
\thanks{X. J. Yang is with Industrial and Operations Engineering, The University of Michigan, Ann Arbor, 500 S State St, Ann Arbor, MI 48109 USA (e-mail: xijyang@umich.edu).}
\thanks{F. Zhou is with the Department of Industrial and Manufacturing, Systems Engineering, The University of Michigan, Dearborn, 4901 Evergreen Rd. Dearborn, MI 48128 USA (e-mail: fezhou@umich.edu).}
\thanks{Manuscript received October 9, 2021; revised xxx 26, 2021.}}

%
%

\markboth{}%
{Zhang \MakeLowercase{\textit{et al.}}: Disengagement Cause-and-Effect Relationships Extraction Using an NLP Pipeline}
%



\maketitle

\begin{abstract}
The advancement in machine learning and artificial intelligence is promoting the testing and deployment of autonomous vehicles (AVs) on public roads. The California Department of Motor Vehicles (CA DMV) has launched the Autonomous Vehicle Tester Program, which collects and releases reports related to Autonomous Vehicle Disengagement (AVD) from autonomous driving. Understanding the causes of AVD is critical to improving the safety and stability of the AV system and provide guidance for AV testing and deployment. In this work, a scalable end-to-end pipeline is constructed to collect, process, model, and analyze the disengagement reports released from 2014 to 2020 using natural language processing deep transfer learning. The analysis of disengagement data using taxonomy, visualization and statistical tests revealed the trends of AV testing, categorized cause frequency, and significant relationships between causes and effects of AVD. We found that (1) manufacturers tested AVs intensively during the Spring and/or Winter, (2) test drivers initiated more than 80\% of the disengagement while more than 75\% of the disengagement were led by errors in perception, localization \& mapping, planning and control of the AV system itself, and (3) there was a significant relationship between the initiator of AVD and the cause category. This study serves as a successful practice of deep transfer learning using pre-trained models and generates a consolidated disengagement database allowing further investigation for other researchers.
\end{abstract}

\begin{IEEEkeywords}
Autonomous vehicles, disengagement, cause-and-effect extraction, natural language processing, deep transfer learning.
\end{IEEEkeywords}

%
\IEEEpeerreviewmaketitle

\section{Introduction}
%
%
%
%
\IEEEPARstart{T}he advancement of machine learning and artificial intelligence is bringing Autonomous Vehicles (AVs) closer to the roads with potential benefits to save people's lives, smooth traffic flow, improve transportation efficiency, reduce energy consumption, and so on \cite{ayoub2021modeling}. According to the definition from the Society of Automotive Engineers (SAE) \cite{sae2018taxonomy}, there are six levels of diving automation from Level 0 (No Driving Automation) to Level 5 (Full Driving Automation). While manufacturers are making the effort to deliver Level 5 AVs in the future, most AVs driving or testing on public roads nowadays are SAE Level 2 (partial automation), Level 3 (conditional automation), and Level 4 (high automation in geofenced areas) vehicles. Particularly, SAE Level 3 automation does not require the human driver to monitor the driving environment \cite{sae2018taxonomy}, which allows the driver to shift attentional resources to non-driving-related tasks \cite{merat2019out}. Even though the driver is required to be ready to take over control whenever requested, s/he may stay out of the control loop for a prolonged period, thus not monitoring the driving situation \cite{gold2018modeling, zhou2019takeover,du2020examining,du2020predicting,zhou2021using}. The transition of control from the AV to the human driver, also known as the Autonomous Vehicle Disengagement (AVD), plays a vital role in the Level 3 AV testing and deployment. Understanding the cause-and-effect relationships of the AVD not only helps to evaluate the safety and performance of the current AV systems, but also provides guidelines on redesign of the AV systems for improvement and the future regulations for AVs.

Generally speaking, investigations into AVD can be divided into two categories, 1) experimental studies in takeover transitions in driving simulators and 2) analysis of naturalistic studies. In the first category, many researchers examined the influences of critical factors on takeover performance, such as takeover lead time (e.g., \cite{du2020evaluating}), non-driving related tasks (e.g., \cite{zeeb2017steering,wandtner2018effects}), traffic densities (e.g., \cite{du2020evaluating,gold2016taking}), emotions \cite{du2020examining}, and drivers' characteristics (e.g., \cite{zeeb2017steering,clark2017age}). This type of studies linked drivers' behavior to well-controlled takeover transitions and examined how critical human factors influenced takeover performance. For example, Du et al. \cite{du2020examining} found that positive emotions (valence) led to better takeover quality, while arousal did not influence takeover time. 
However, due to the gap between driving simulation and naturalistic driving, their findings may only be transferred to real world in a specific scope with limitations. For example, it was found that participants' perceived risk in driving simulator was lower compared to that in naturalistic driving \cite{carsten2011driving}.  

The second category of studies focused on analyzing takeover transitions in naturalistic driving, particularly by examining the disengagement reports released by the California Department of Motor Vehicles (CA DMV) using statistical methods (e.g., \cite{FAVARO2018136,2020explore,WANG201944}). The CA DMV disengagement dataset consisted of specific disengagement reports in naturalistic driving submitted by testing companies from 2014 to 2020, covering important data fields, such as location, initiator, description of disengagement, which was perfect for analyzing the cause-and-effect relationships of AVD. For example, Boggs et al. \cite{2020explore} explored five W questions (i.e., who, what, when, where, and why) of AVD using CA DMV disengagement reports from September 2014 to November 2018. They found that 1) the test operators were more likely to initiate the disengagement on streets and roads than on freeways and interstates, 2) the major causes to system-initiated disengagement were software, hardware, and planning issues, and 3) system-initiated disengagement marginally increased compared to operator-initiated disengagement despite the advancement in AV technologies between 2014 and 2018. However, in order to conduct such studies, laborious manual processing of the reports is needed.


In consideration of the pros and cons of the two types of studies, in this paper, we built a natural language processing (NLP) pipeline to extract cause-and-effect relationships from disengagement reports of the most updated CA DMV dataset. This NLP pipeline promised 1) to process various formats of disengagement data at scale, 2) to analyze two types of disengagement on future incoming reports, and 3) to provide implications on AV testing and deployment without human manual work. To achieve those goals, the NLP pipeline utilized deep transfer learning, which improved the learning of a new task, i.e., cause-and-effect relationship extraction of ADV ``through the transfer of knowledge from related natural language understanding tasks that have already been learned" \cite{torrey2010transfer}. By doing so, our deep learning model achieved human-level performance with only a small portion of manual labeled data. 

As a summary, the contributions of this study are: 
\begin{itemize}
  \item We proposed an NLP pipeline to process and analyze AVD reports at scale with satisfactory performance.
  \item We built an NLP deep learning model that was first pre-trained on a large-scale natural language corpus (e.g., Wikipedia, Google News), then fine-tuned on a task-specific dataset (SemEval-2010 Task 8 \cite{hendrickx2019semeval2010}) extended by Li et al. \cite{Li_2021} to include embedded causality \cite{mostafazadeh-etal-2016-caters}, and finally post-trained on the CA DMV dataset.
  \item We developed a taxonomy of the causes and effects of AVD with a focus on how the AV system worked and interacted with other factors.
  \item We identified the significant relationships between causes and initiators of AVD and analyzed contributions of causes of two types of AVD, which provided valuable insights into the current issues existing in the AV system and potential improvements for manufacturers to improve AV safety.
\end{itemize}

\section{Related Work}
\subsection{Experiment Studies in Driving Simulators}
Many previous studies conducted driving simulation to investigate the influence of individual factors on takeover performance in takeover transitions and analyzed driver behaviors under various controlled scenarios related to AVD. 
For example, Du et al. \cite{du2020evaluating} examined the effects of three factors on takeover performance, including drivers' cognitive workload, takeover time budget, and traffic density. They found that takeover performance was worse when the participants had a high level of cognitive workload, the time budget was small, and the oncoming traffic density was heavy. Wandtner et al. \cite{wandtner2018effects} investigated the effects of non-driving related tasks on takeover performance and found that visual-manual tasks significantly worsened takeover performance compared to auditory-vocal tasks. Clark and Feng \cite{clark2017age} found that younger (age 18-35, $n=17$) and older drivers (age 62-81, $n=18$) showed different behaviors in simulated automated driving, such as different preferences for non-driving related tasks, different takeover reaction time, and significant impacts on driving performance.
Gold et al. \cite{gold2018modeling} designed takeover experiments, identified related variables, and developed regression models to evaluate takeover performance. Their study revealed the effects of key factors, such as time budget, traffic density, and repetition, on takeover performances.    
Markkula et al. \cite{markkula2018models} created simulation-ready human behavior models to reproduce qualitative patterns of important scenarios like ``an AV handing over control to a human driver in a critical rear-end situation''. With driving simulations, their models allowed optimization of AV impacts on safety. 

These experimental studies based on driving simulation discovered various findings of takeover transitions, human factors, and potential causes of AVD. And the findings may be transferred to open roads to help improve the transition process of control from the AV system to the human driver. Even though Eriksson et al. \cite{eriksson2017transition} have found strong positive correlation between the driving simulation and on-road conditions for control transitions in terms of workload, perceived usefulness and satisfaction, on-road validation is still necessary for those results discovered through driving simulation. Not only because the AV system used in driving simulation cannot reproduce the driving experience provided by various AV systems developed by different manufacturers, but also only a small number of individual factors were considered in the models developed on the experimental data from simulators \cite{gold2018modeling,eriksson2017transition}. Furthermore, the number of participants for driving simulation tasks was mostly less than 100 and the number of records of experimental data was mostly less than 1000 \cite{gold2018modeling,eriksson2017transition, gold2013take,gold2014influence,radlmayr2014traffic,korber2016influence,du2020predicting}. The number of the participants and the amount of the data may limit the generalizability and scalability of those studies.  
\begin{figure*}
    \centering
    \includegraphics[width=0.8\textwidth]{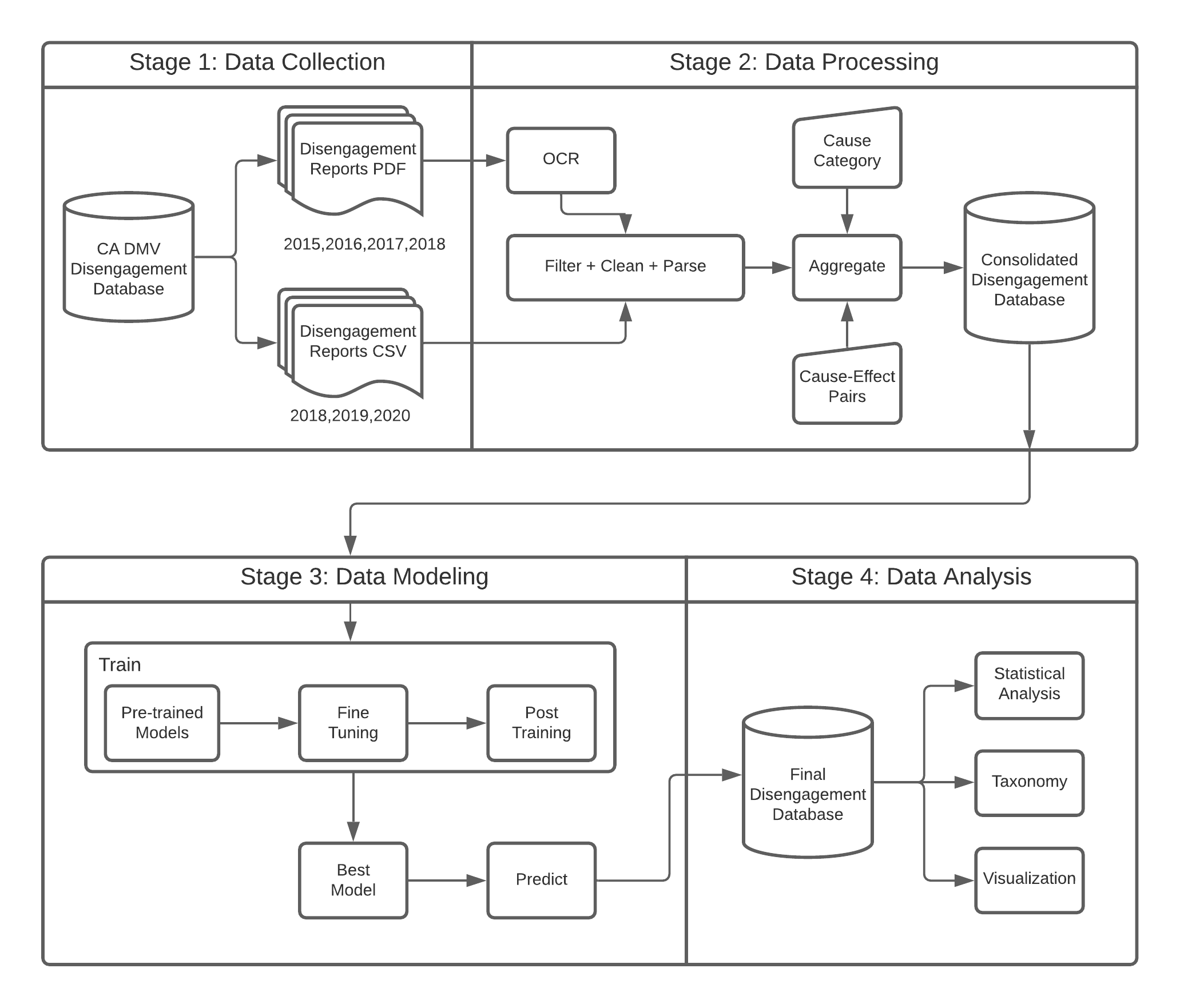}\hfill
    \caption{Overview of the data pipeline from Stage 1 to Stage 4}
    \label{fig:pipeline}
\end{figure*}
\subsection{Naturalistic studies}
With the availability of the CA DMV dataset, many researchers focused on the exploration of the disengagement reports using traditional statistical and machine learning models. For example, Favarò et al. \cite{FAVARO2018136} analyzed the contributory factors, disengagement frequencies of AVD with taxonomy and statistical visualization of disengagement overview and trends. Boggs et al. \cite{2020explore} used logistic regression to identify and quantify who, what/why, where, and when (5 Ws) of AVD, and they found that 1) it was more likely to have human operator-initiated disengagement than system-initiated, 2) the most frequent causes of disengagement of system-initiated were planning and software/hardware discrepancies, 3) disengagement was more likely to happen on local roads than on expressways, and 4) regardless of system improvement, the likelihood of system-initiated disengagement increased marginally compared to human operator-initiated. Wang et al. \cite{WANG201944} used multiple statistical modeling approaches and classification trees to quantitatively investigate the underlying causes of AVD, and found that lacking a certain number of sensors significantly induced AVD. However, the researchers in these studies had to manually process the reports in various formats, which was time-consuming and laborious. Thus, these studies lacked the scalability to handle large-scale data and cannot be applied on new incoming disengagement reports to benefit future analysis. On the other hand, other researchers applied natural language processing techniques to process the text data automatically. For example, Alambeigi et al. \cite{alambeigi2020crash} used probabilistic topic modeling to identify themes from the CA DMV crash reports. Their findings emphasized the safety concerns with transitions of AV system to human control. However, the number of crash reports was much smaller (167 reports in \cite{alambeigi2020crash}), compared to the disengagement reports.  Banerjee et al. \cite{2018handsoff} applied the data pipeline method to process and analyze data from the system's perspective with over a million miles of cumulative autonomous miles, 5,328 disengagement reports, and 42 crash reports from 2014 and 2017. They found that 1) AVs were 15 - 4000 times worse than human drivers in terms of accidents per cumulative mile driven, 2) perception, decision, and control discrepancies resulting from the AVs' machine-learning-based system were the primary causes of AVD, and 3) human operators of AVs had to stay as alert as drivers of manual vehicles. Such a method provide promises to automatically process the data systematically. However, the NLP model was based on keywords matching and voting, which could end up with many unknown instances when assigning the causes to a specific predefined category.


\section{NLP Pipeline}

A scalable end-to-end pipeline (Fig. \ref{fig:pipeline}) was constructed to collect, process, model, and analyze the disengagement reports with high efficiency and accuracy. The pipeline consisted of four stages. 
At Stage 1, it collected multi-format disengagement reports from the CA DMV disengagement database and classified them into corresponding years. 
At Stage 2, it used Optical Character Recognition (OCR) to extract information from the PDF format reports and exported to CSV files. Then those CSV files were filtered, cleaned, parsed, and finally labeled by human workers. 
At Stage 3, we proposed an NLP deep learning model based on ELECTRA (Efficiently Learning an Encoder that Classifies Token Replacements Accurately) \cite{clark2020electra} using transfer learning. In order to compare with other popular NLP deep learning models, BERT \cite{devlin2019bert}, DistilBERT \cite{sanh2020distilbert}, and XLNet \cite{yang2020xlnet} were also included. 
At Stage 4, it analyzed the disengagement database by creating a taxonomy, summarizing results in visualization and statistical analysis. 

\subsection{Data Collection}
In September 2014, the CV DMV initiated the Autonomous Vehicle Tester Program which allowed permit-holding manufacturers to test AVs with a test operator in the driver seat on public infrastructure \cite{cadmv}. The program also required permit holders to track and submit disengagement reports when ``their vehicles need to disengage from the autonomous mode during tests"  and collision reports for ``every collision involving one of their vehicles" \cite{cadmv}. Since the establishment of the program, the CA DMV has been releasing disengagement reports and collision reports to the public on a yearly basis. These reports consisting of thorough and specific raw data are ideal data sources for investigating AVD.

The raw disengagement data were retrieved from the CA DMV's disengagement database, which included the disengagement reports collected from manufacturers participating in the Autonomous Vehicle Tester Program on a yearly basis. For this study, the disengagement reports released from 2014 to 2020 were used as the initial data for processing and modeling. However, the proposed method can be updated easily with new data.
\subsection{Data processing}
\textbf{Cleaning:} The format requirement of the disengagement reports has changed over time, including photoshopped non-regulated formats across different manufactures between 2014 and 2017, standard PDF template in part of 2018, and consolidated CSV files between 2018 and 2020. For the latest CVS format, it combined all the records together and released two final CSV files - Autonomous Vehicle Disengagement Reports and Autonomous Mileage Reports, which were easy to process and distribute to the public. For each record, the final Autonomous Vehicle Disengagement Reports contained 9 fields. In order to resolve these differences across different formats, we attempted to collect all the information in the previous years (2014 - 2018) based on the latest format (2018 - 2020). An OCR pipeline based on OpenCV \cite{opencv_library}, Tesseract \cite{Tesseract} and PyImageSearch \cite{pyimagesearch} was built to extract texts from the PDF files, compile them, and export them to CSV files. As the PDF files submitted by various manufacturers were scanned, they were subject to random scaling, rotation, and skew. To fix this problem, a standard template was used as the reference to adjust other scanned disengagement reports to generate the de-skewed versions of images. The bounding boxes were then identified and the text within them were extracted. Next, raw texts were cleaned, filtered, and finally exported as CSV files. Finally, we collected the disengagement reports in the last 7 years (2014 - 2020) and collected and inspected, manufacturer, date, initiator, location and description were selected as the fields for the consolidated disengagement database for further analysis. Table~\ref{tab:sample} shows the sample data and format in the database. N/A was introduced as the placeholder for missing values to ensure the integrity of the database.


\begin{table*}
  \centering
  \small
  \caption{Sample data and format for collected reports.}
\begin{tabularx}{\textwidth}{llllX}

\hline\hline
Manufacturer & Date & Initiator & Location & Description \\ 
\hline
EasyMile & 11/30/2020 & AV System & Street & A collision hazard in the environment ahead was detected by the software, which triggered an emergency stop \\

Apple & 06/19/2019 & AV System & Street & Motion planning timed out \\

Uber & 03/01/2018 & Test Driver & Street & Precautionary Takeover or Operator Discretion \\

Waymo & 09/01/2017 & N/A & Highway & Disengage for a software discrepancy \\

Tesla & 10/15/2016 & AV System & Freeway & Follower Output Invalid \\

Volkswagen & 06/12/2015 & N/A & N/A & Planner not ready \\ 
\hline
\end{tabularx}
\label{tab:sample}
\end{table*}

\textbf{Filtering:} The collected data with various formats required further data preprocessing. The disengagement reports with cause description less than five words were removed to reduce the noise existed in the raw data. In addition, as suggested by Boggs et al. \cite{BOGGS2020105354}, ``Apple and Uber lacked a variation among human-initiated disengagements" in terms of cause description and those disengagement reports replicated the same information hundreds of thousands of times. Therefore, those records were excluded as well. 

\begin{figure}
    \centering
    \includegraphics[width=1\columnwidth]{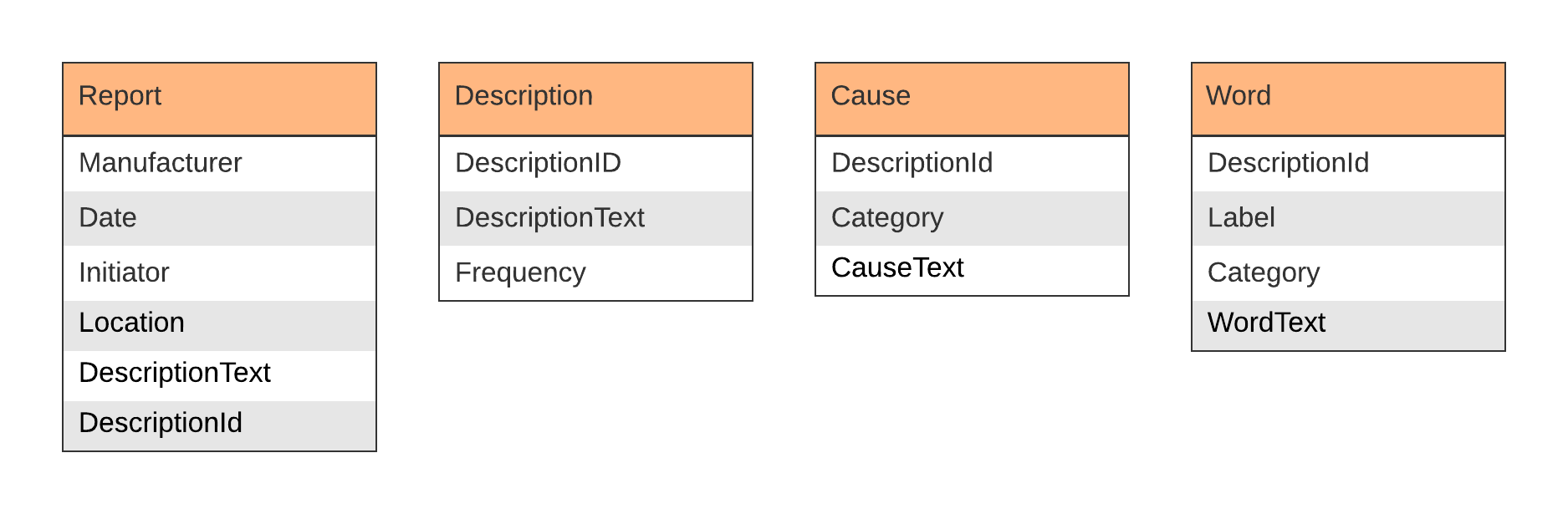}
    \caption{Overview of entities in the database}
    \label{fig:database}
\end{figure}
\begin{figure}
    \centering
    \includegraphics[width=1\columnwidth]{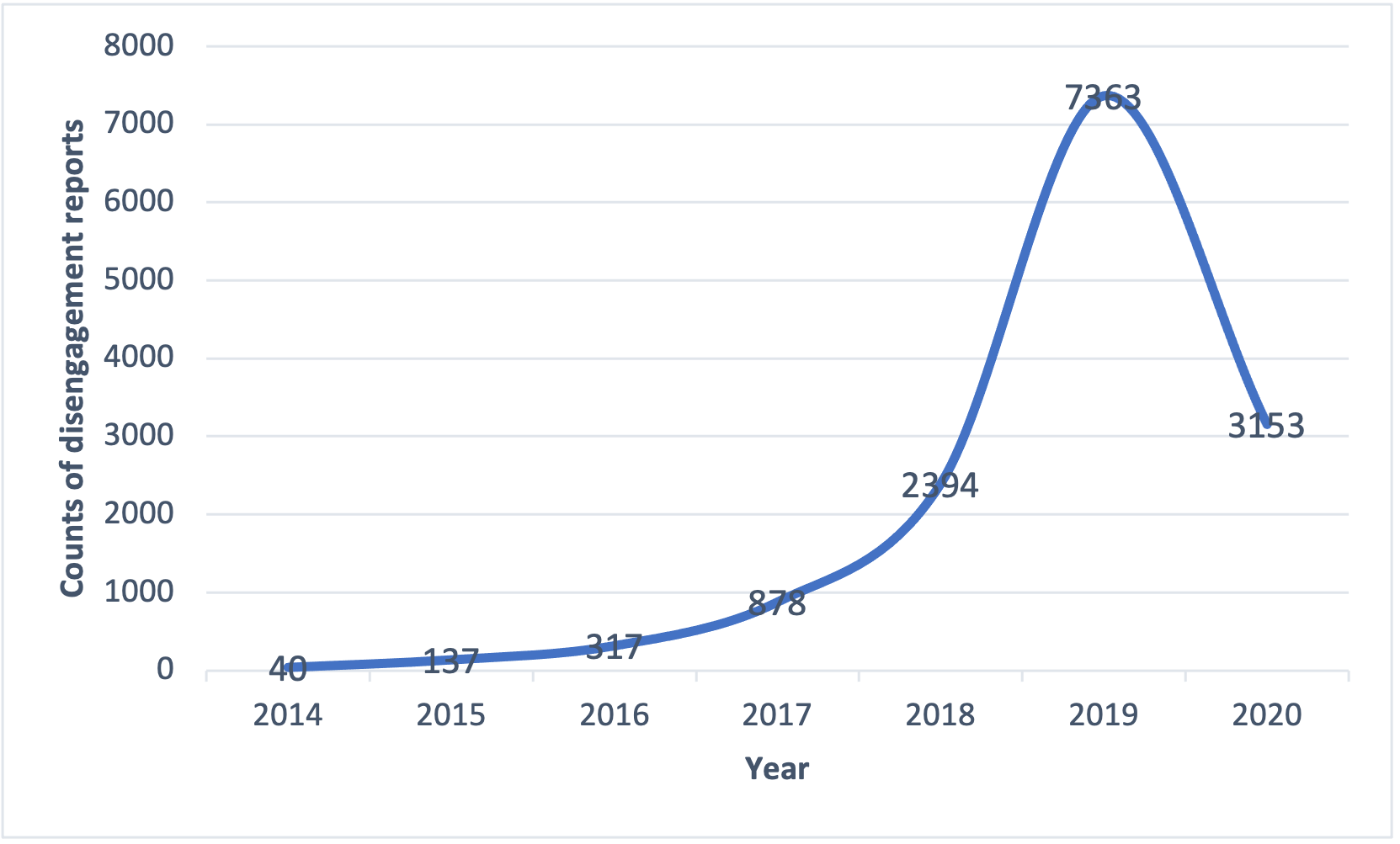}
    \caption{The number of filtered disengagement reports of each year in the consolidated database.}
    \label{fig:count_year}
\end{figure}

With all these preprocessing steps, a consolidated database (Fig. \ref{fig:database}) was generated. It contained information of four entities - report, description, cause, and word. The four entities were weaved together logically - reports had descriptions, causes existed in descriptions, and causes were made of words. Fig. \ref{fig:count_year} shows the number of filtered disengagement reports of each year submitted by manufacturers in the database with a total number of 14, 282 reports. 

\textbf{Human Insights:}  Two types of human insights were required for this study to better understand the dataset: cause-and-effect relationship labeling and cause categorization. The former was used to identify the causes and effects from the disengagement reports for training the NLP deep learning models and the latter contributed to classifying the causes into proper categories. Three student workers, who had at least 3 months experience in this field, provided the human insights.

For cause-and-effect relationships, both simple causality and embedded causality were considered in this study. Following the standard cause-and-effect tagging format  \cite{Li_2021,mostafazadeh-etal-2016-caters}, the Inside, Outside, Beginning (IOB) notations were used to label the tokens. To be more specific, suffix C - Cause, E - Effect and CE - Embedded Cause were added to the IOB notations. For example, B-C represents the beginning of a cause token while I-E represents the interior of an effect token. Based on human insights, each word in the description of disengagement was tagged with labels and categories and an example is shown in Table \ref{tab:bio}.

For the cause category, three main categories (AV System, Human Factors, Environmental Factors \& Others) and nine subcategories (Perception, Localization \& Mapping, Planning, Control, System General, AV Driver, Other Driver \& Vehicle, Environment, Other) were derived from the conceptual organization highlighting trust development produced by Schaefer et al. \cite{2016metafactors}, the AV hierarchical control structure drawn by Banerjee et al. \cite{2018handsoff}, and the taxonomy developed by Boggs et al. \cite{2020explore}. 

\begin{table}
\renewcommand{\arraystretch}{1.3}
\centering
\caption{Sample sentence for IOB labelling and cause categorization. (B-E represents the beginning of an effect, I-E represents the interior of an effect, B-C represents the beginning of a cause, I-C represents the interior of a cause, and O represents others. 2 means the cause belongs to the planning category).}
  \centering
  \small
\begin{tabular}{lll}
\hline\hline
\textbf{Words}                     & \textbf{Label}                    & \textbf{Category}                \\ \hline
driver                             & B-E                               &                                  \\ 
disengagement                      & I-E                               &                                  \\ 
due                                & O                                 &                                  \\ 
to & O &  \\ 
planning                           & B-C                               & 2 - planning                                \\ 
discrepancy                        & I-C                               & 2 - planning                                \\ 
in                                 & O                                 &                                  \\ 
the                                & O                                 &                                  \\ 
determination                      & O                                 &                                  \\ 
of                                 & O                                 &                                  \\ 
autonomous                         & O                                 &                                  \\ 
vehicle                            & O                                 &                                  \\ 
speed                              & O                                 &                                  \\ \hline
\end{tabular}
\label{tab:bio}
\end{table}

While different workers provided their insights, it was necessary to aggregate those insights and to determine the ground truth. The framework CrowdTruth \cite{CrowdTruth2} was used for ground truth aggregation. It not only selected the most reliable labels among workers but also offered useful metrics, including Worker Quality Score (WQS) and Annotation Quality Score (AQS) to evaluate the performance of workers as well as the quality of their annotations. As shown in Table \ref{tab:aqs}, the WQS and AQS of the ground truth aggregation were higher than 0.9, which provided convincing support for the quality of the human annotation of all the disengagement reports.

\begin{table}
\renewcommand{\arraystretch}{1.3}
\centering
\caption{Annotation quality score for eight labels and worker quality score for three workers.}
\begin{tabular}{llcc}
\hline\hline
\textbf{Label} & \textbf{AQS}    & \textbf{Worker ID} & \textbf{WQS}\\ \hline
O              & 0.9942          & 0 & 0.9802\\ 
B-C            & 0.9257          & 1 & 0.9751\\ 
I-C            & 0.9352          & 2 & 0.9845\\ 
B-E            & 0.9461          \\ 
I-E            & 0.9184          \\ 
B-CE           & 0.9042          \\ 
I-CE           & 0.9331          \\ \hline
\end{tabular}
\label{tab:aqs}
\end{table}


\subsection{Cause and Effect Modeling}
We proposed an NLP deep learning model based on ELECTRA \cite{clark2020electra} using transfer learning. First, ELECTRA pre-trained on Wikipedia and BooksCorpus consisting of 3.3 Billion tokens was imported. ELECTRA is a novel NLP deep learning model that outperformed multiple existing techniques, including BERT \cite{devlin2019bert}, DistilBERT \cite{sanh2020distilbert}, and XLNet \cite{yang2020xlnet}, on numerous natural language understanding benchmark with the same computational resources.
Second, in order to gain task-specific knowledge for extracting cause-and-effect relationships, ELECTRA was fine-tuned on the SemEval-2010 Task 8 \cite{hendrickx2019semeval2010}).The SemEval-2010 Task 8 is a benchmark for multi-way classification of semantic relations between pairs of nominals. In this dataset, we used the samples with annotated cause-and-effect relationships to fine-tune the ELECTRA model. Third, ELECTRA was post-trained on our consolidated disengagement dataset created in Stage 2 of the data pipeline to further improve the performance of the models. In order to compare with other popular NLP deep learning models, BERT \cite{devlin2019bert}, DistilBERT \cite{sanh2020distilbert}, and XLNet \cite{yang2020xlnet} were also included. These models were evaluated based on weighted F-1 scores and the cost of computational resources to select the best model. Among them, weighted F-1 scores calculated the average F-1 scores among labels weighted by their support to handle label imbalance existed in the training data and the cost of computational resources was measured by the time needed to train the models with the same configurations of the computer. 

During the transfer learning process, the target task was defined with two different approaches to maximize the benefits of transfer learning. The first approach utilized two same-type pre-trained models with different heads for different purposes - Model One was fine-tuned for cause-and-effect relationship extraction with a token classification head and Model Two was fine-tuned for cause category classification with a sequence classification head. The two models were chained together, so that the extracted causes from Model One were fed into Model Two directly to obtain their categories. This approach allowed Model One to benefit from post-training but dropped the sentence context for Model Two. The second approach only involved one end-to-end model. The tagged labels and categories were combined (Table \ref{tab:label} shows how the combination works) to satisfy the requirements of the end-to-end model. Compared with the first approach, the second approach was more computationally efficient and enabled the usage of sentence context for category prediction.

\begin{table}
  \centering
  \small
\caption{How labels and categories were combined.}
\begin{tabular}{llll}
\hline\hline
Words         & Label      & Category & Combined Label \\ \hline
driver        & B-E        &          & B-E            \\
disengagement & I-E        &          & I-E            \\ 
due           & O &          & O              \\ 
to            & O          &          & O              \\
planning      & B-C        & 2 - planning        & B-C-2          \\ 
discrepancy   & I-C        & 2 - planning        & I-C-2          \\ \hline
\end{tabular}
\label{tab:label}
\end{table}
\subsection{Statistical Summary and Visualization}
We analyzed the disengagement database by creating a taxonomy at Stage 4, summarizing the results with statistical analysis and visualization. In terms of the taxonomy, the AVD causes were classified into categories and subcategories. The frequency and distribution of the causes in different categories revealed insights, such as ``in which stage, the AV system failed to execute tasks most frequently" and ``which environmental factors caused AVD most often". Various visualizations offered more direct and intuitive insights. For example, the word cloud gave an impression of those hot words used frequently to describe AVD while the time-series graph showed how the cases of AVD changed over the past seven years and demonstrated patterns which deserved further investigation. In addition, the statistical analysis provided more quantitative findings of significant relationships between variables with statistical tests. 

\section{RESULTS}
\subsection{Cause-Effect Extraction}
For the first approach described in Section \emph{Cause and Effect Modeling}, following the best practice of transfer learning \cite{loshchilov2019decoupled}, the setting of the four pre-trained deep learning models were shown as follows: Fine-tuned using AdamW as the optimizer with an initial learning rate of $5\mathrm{e}{-5}$, and a learning rate scheduler decreasing the learning rate linearly in 15 epochs.  To investigate the generalization of the models to new data, five-fold cross-validation was applied. The whole labeled dataset was partitioned evenly into 5 complementary subsets based on the distribution of the labels. Each subset was used for testing once while the rest four subsets were used for training. In this manner, the whole dataset was fully utilized and the scores of the models were averaged over five testing subsets to generate the final balanced score.

As shown in Table \ref{tab:ce}, among the four pre-trained models, ELECTRA achieved the highest weighted F-1 score with relatively low computation resources. Furthermore, post-training significantly improved the performance of the best model at the cost of longer training time. The pre-trained model with fine-tuning and post-training achieved the best performance.

\begin{table}
  \centering
\caption{Weighted F-1 score and training time for cause-and-effect relationships extraction.}
\begin{tabular}{lcc}
\hline\hline
Model                                                       & Weighted F-1                         & Training Time                       \\ \hline
BERT + Fine-tuning                                                    & 0.76                                 & 17min 39s                                 \\ 
XLNET + Fine-tuning                                                   & 0.78                                 & 24min 52s                                 \\ 
DistillBERT + Fine-tuning                                             & 0.75                                 & \textbf{10min 30s}                                 \\
ELECTRA + Fine-tuning                                                 & 0.82                                 & 17min 46s                                 \\ 
ELECTRA + Post-training                                               & 0.69                                 & 12min 36s                                 \\ 
ELECTRA + Fine-tuning + Post-training &\textbf{0.90}& 28min 14s\\ \hline
\end{tabular}
\label{tab:ce}
\end{table}

\subsection{End-to-End Token Classification}
For the second approach described in Section \emph{Cause and Effect Modeling}, the training setup for this approach was the same as the first approach. Since more specific labels were used for the end-to-end model, the difficulty and complexity for prediction increased significantly. However, as shown in Table \ref{tab:e2e}, the pre-trained models still achieved relatively good performance. If the results in Table \ref{tab:ce} was compared to those in Table \ref{tab:e2e}, it is satisfactory to find that the weighted F-1 score of ELECTRA model with fine-tuning only dropped from 82\% to 75\% while the complexity of token classification task scaled up to almost 8 times from 7 tags (O, B-C, I-C, B-E, I-E, B-CE, I-CE) to 55 tags (O, B-C-0, ... , I-C-0, ... , B-E-0, ... , I-E-8, ... , B-CE-8, ..., I-CE-8).

\begin{table}
  \centering
\caption{Weighted F-1 score and training time for end-to-end token classification.}
\begin{tabular}{lcc}
\hline\hline
\textbf{Model}                                        & \textbf{Weighted F-1}                         & \textbf{Training Time}                         \\ \hline
BERT + Fine-tuning                                    & 0.72                                 & 17min 59s                                 \\ 
XLNET + Fine-tuning                                   & 0.74                                 & 24min 57s                                 \\ 
DistillBERT + Fine-tuning                             & 0.71                                 & \textbf{10min 37s}                                 \\ 
ELECTRA + Fine-tuning & \textbf{0.75} & 18min 12s \\ \hline
\end{tabular}
\label{tab:e2e}
\end{table}

\begin{figure*}[h]
    \centering
    \includegraphics[width=0.7\textwidth]{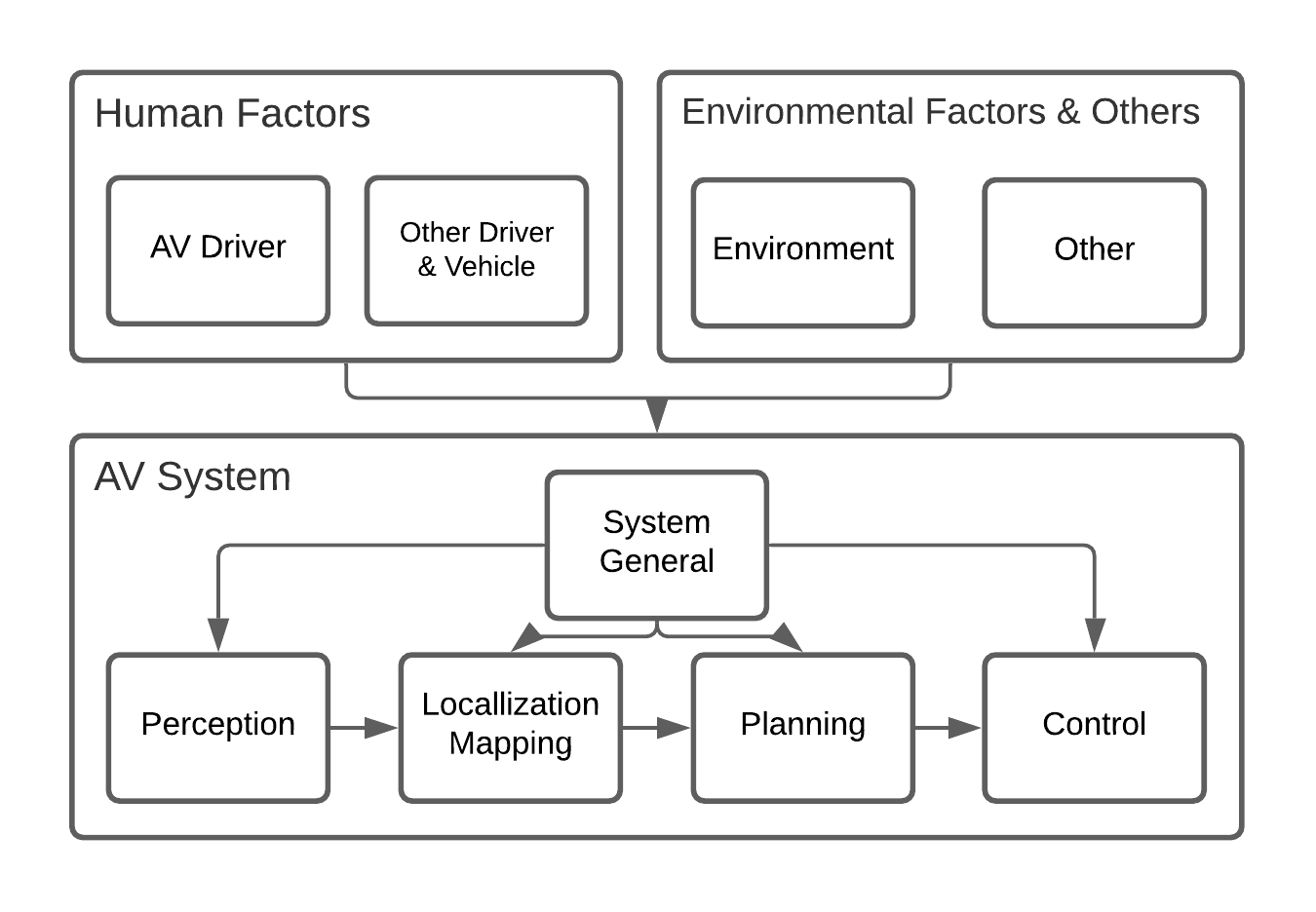}
    \caption{Overview of the taxonomy, which has three main categories of causes, including human factors, AV system, environmental factors and others. Each main category also has their own subcategories as shown in the figure.}
    \label{fig:taxonomy}
\end{figure*}

\subsection{Taxonomy}
Schaefer et al. \cite{2016metafactors} used a meta-analysis method to identify three major factors influencing the trust development in automation, including human factors, system factors, and environment factors. This potentially caused the operators to initiate the disengagement due to a lack of trust \cite{ayoub2021modeling}. Similarly, we adopted this method to categorize the causes of the disengagement. Furthermore, the failure of different components - perception, planning, control \cite{machines5010006} in the AV system also led to disengagement. Such failures could be caused by the environment and the system itself. Thus, a taxonomy (Fig. \ref{fig:taxonomy}) covering both human driver trust and AV system failures, and environmental factors was generated. This taxonomy served as the basis of the categories and subcategories for causes, which made our analysis more intuitive and efficient. Table \ref{tab:category} shows the number of unique causes in each category.

\begin{table}
\centering
\caption{The number of unique causes in different categories.}
\begin{tabular}{lc}
\hline\hline
Cause Category     & Count \\ \hline
0 - perception              & 52             \\ 
1 - localization \& mapping & 43             \\ 
2 - planning                & 52    \\ 
3 - control                 & 59             \\ 
4 - AV driver               & 41             \\ 
5 - other driver \& vehicle & 60             \\ 
6 - environment             & 33             \\ 
7 - system general          & 29             \\ 
8 - others                   & 8              \\ 
Total                       &377                \\ \hline
\end{tabular}
\label{tab:category}
\end{table}

\begin{table}[]
\caption{The contingency table for initiator and cause category.}
\label{tab:contingency}
\begin{tabular}{lcc}
\hline\hline
\multirow{2}{*}{Cause Categories} & \multicolumn{2}{c}{Initiator} \\
 & AV Systems & Test Operators \\\hline
AV System & 1703 & 5493 \\
Human Factors & 1 & 1871 \\
Environmental Factors and Others & 46 & 397 \\\hline
\end{tabular}%
\end{table}

\begin{table}[]
\caption{The contingency table for initiator and sub cause category.}
\label{tab:contingency_sub}
\resizebox{\columnwidth}{!}{%
\begin{tabular}{llcc}
\hline
\multirow{2}{*}{Cause Category} & \multirow{2}{*}{Cause Subcategories} & \multicolumn{2}{c}{Initiator} \\
 &  & AV Systems & Test Operators \\ \hline
\multirow{4}{*}{AV System} & 0 - perception & 322 & 998 \\
 & 1 - localization \& mapping & 106 & 221 \\
 & 2 - planning & 775 & 1423 \\
 & 3 - control & 71 & 2291 \\
\multirow{2}{*}{Human Factors} & 4 - AV driver & 0 & 1534 \\
 & 5 - other driver \& vehicle & 1 & 337 \\
\multirow{3}{*}{\begin{tabular}[c]{@{}l@{}}Environmental \\ Factors and Others\end{tabular}} & 6 - environment & 38 & 185 \\
 & 7 - system general & 429 & 560 \\
 & 8 - others & 8 & 212 \\ \hline
\end{tabular}%
}
\end{table}
\begin{figure}[h]
    \centering
    \includegraphics[width=\linewidth]{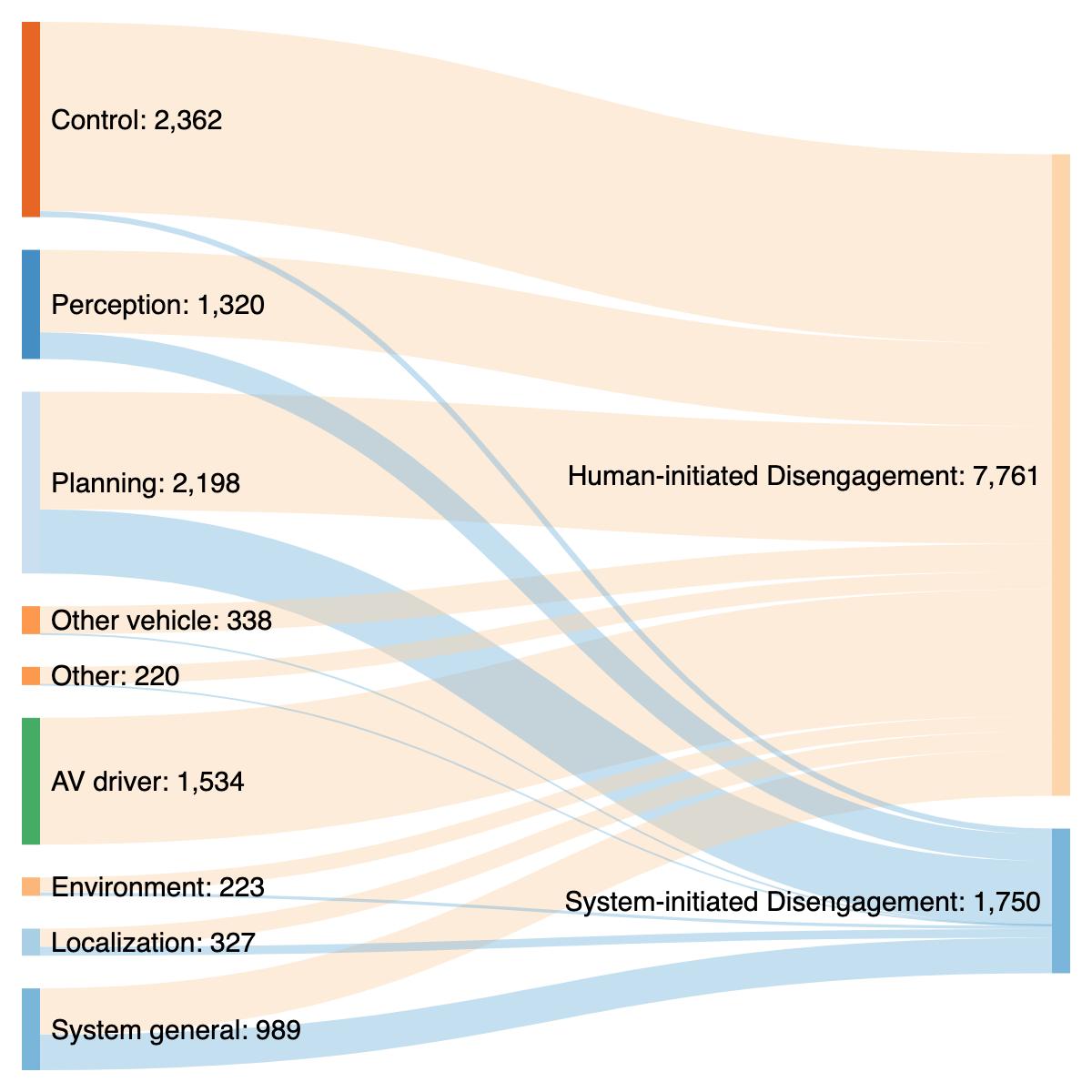}
    \caption{Sankey chart for cause category contribution to two types of disengagement.}
    \label{fig:sankey}
\end{figure}
\subsection{System-Initiated vs. Human-Initiated Disengagement}
The pipeline can also conduct multiple statistical tests to answer questions related to two types of disengagement, such as ``Was there a significant relationship between the initiator and the cause category?" or ``How did the different cause category contribute to the two types of disengagement?". Fig \ref{fig:sankey} shows how the different cause categories contributed to two types of disengagement in the consolidated database. 
Table \ref{tab:contingency} and Table \ref{tab:contingency_sub} were the contingency tables for initiators between the categories and the subcategories. Table \ref{tab:contingency} shows that for AVD initiated by AV systems and test drivers, most causes came from the AV systems themselves. 
Table \ref{tab:contingency_sub} provided a more detailed insight suggesting that for AVD initiated by the AV systems, the planning stage was the most unreliable stage in the AV systems, while for AVD initiated by the test operators, the majority were caused by either the control stage of the AV systems or the discomfort felt by the test operators. In addition, Chi-Square tests for independence were conducted on the two contingency tables. There was a significant relationship between the initiators of AVD and the main categories, ${\chi}^2 (2, 9511) = 571.53, p < 0.001$.
And the frequency of the causes in the subcategories differed significantly by the initiators as well, ${\chi}^2 (8, 9511) = 1726.13, p < 0.001$.
According to the disengagement reports released by CA DMV, the initiator of disengagement was either the AV System when it failed to execute due to technical issues and thus requested the test operator to take over control, or the test operator when he/she felt uncomfortable with or did not trust the AV system and thus took over control proactively.


\subsection{Visualization}
\begin{figure}[]
    \centering
    \begin{minipage}[b]{0.48\textwidth}
    \centering
    \makebox[\textwidth][c]{\includegraphics[width=0.99\textwidth]{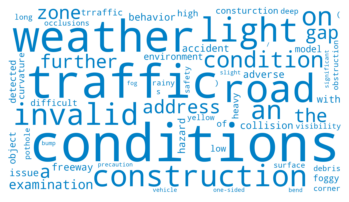}}
    \caption{Sample word cloud for the category environment.}
    \label{fig:wordcloud}
    \end{minipage}
    \begin{minipage}[b]{0.48\textwidth}
    \centering
    \makebox[\textwidth][c]{\includegraphics[width=0.99\textwidth]{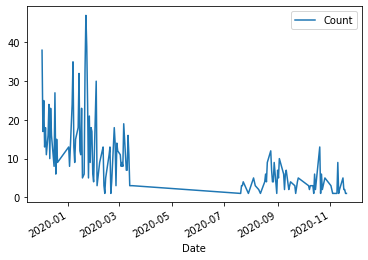}}
    \caption{Sample disengagement time series chart for Toyota in 2020.}
    \label{fig:timeseries}
    \end{minipage}
\end{figure}

The pipeline can produce various visualizations to support further analysis, including but not limited to the word cloud of causes under different categories (e.g., Fig. \ref{fig:wordcloud}), the time series of disengagement reported by manufacturers in specific time range (e.g., Fig. \ref{fig:timeseries}), the multi-series bar chart for causes initiated by the AV system or the human operator, etc.

\begin{figure*}[]
    \centering
    \includegraphics[width=0.9\textwidth]{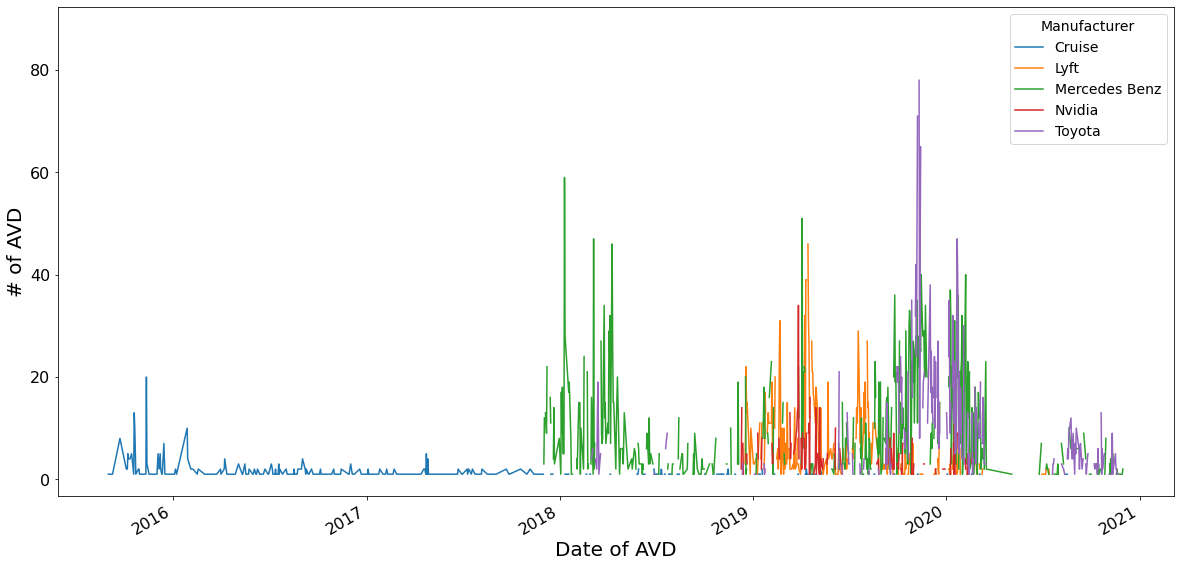}
    \caption{Time series chart for top five manufacturers having the most AVD counts.}
    \label{fig:timeseries_all}
\end{figure*}
Fig. \ref{fig:timeseries_all} demonstrated how cases of reported AVD for top 5 manufacturers changed over time. From the figure, it was easy to find that Cruise started AV testing very early but paused at the end of 2018, while other manufacturers continued to test AVs until early 2020 when COVID-19 happened. After COVID-19 was gradually under control, they resumed the AV testing but had not reached to their full capacity yet by the end of 2020. Some patterns can also be identified from the figure. For instance, those peaks in the figure always happened in the Winter and/or Spring seasons: (1) Cruise - 2015 Winter, 2016 Spring, (2) Lyft - 2019 Spring, (3) Mercedes Benz - 2018 Spring, 2019 Spring, 2019 Winter, 2020 Spring, (4) Nvidia - 2019 Spring, (5) Toyota - 2019 Winter, 2020 Spring. Further researches are needed to investigate whether it was a coincidence or there were reasons for manufacturers to test AVs intensively during these two seasons.

\section{DISCUSSION}
\subsection{Model Performance} 
Our NLP pipeline built on the ELECTRA model was able to extract cause-and-effect relationships with a weighted F-1 score of 0.90, which outperformed other selected deep learning models. Though the training time was longer compared to other models, it was still acceptable within 30 minutes. When we used it for end-to-end token classification, the weighted F-1 score was decreased due to the significant increase in difficulty and complexity of token classification. Nevertheless, the ELECTRA-based model also performed best among the selected NLP deep learning models. Further research should be conducted to further improve the performance of the model. For example, the disengagement scenarios were described in specific domain language and the fine-tuning sample from the SemEval-2010 Task 8 was relatively small. Thus, NLP deep learning models can be further pre-trained in domain language and fine-tuned with a larger sample size of cause-and-effect data. Only three workers provided insights of cause-and-effect relationships and cause categories, which may cause bias. More workers are needed to make the ground truth labels more reliable to improve the performance of the NLP pipeline.

The methods used in this study also have implications for future AVD analysis. As Bimbraw et al. \cite{bimbraw2015autonomous} concluded ``Most cars are expected to be fully autonomous by 2035", before that actually happens, AV testing still requires the presence of human operators which will generate a large amount of disengagement reports.
However, the majority of previous studies heavily relied on manual work and their findings cannot be applied to incoming disengagement reports seamlessly. The weighted F-1 score of this study suggested that with transfer learning, pre-trained models with millions of parameters can be fine-tuned and post-trained to learn the domain knowledge of AVD and the task knowledge of specific analytical tasks, thus achieving satisfying performance close to human workers with much less cost and time. Furthermore, because of the scale of the future disengagement reports, the end-to-end pipeline approach used in this study is both efficient and necessary for the large-scale analysis of AVD that other researchers may be interested in. 

\subsection{Two Types of Disengagement} 
The analysis of disengagement initiators suggests that more than 80\% of the disengagement were initiated by test drivers, who either felt uncomfortable about the maneuver of the AVs or made precautionary takeovers because of insufficient trust. Therefore, additional researches on human trust towards AV and human comfort levels are important to further investigate and explain the manual takeover, which will be a solid step to solve unnecessary disengagement and lay the foundation for the future full automation. The results of various causes related to the AV system also suggest that discrepancies happened in perception, localization \& mapping, planning and control were the primary causes that led to the failures of executing certain tasks by the AV system. Our NLP pipeline not only can summarize the cause-and-effect relationships in the taxonomy as shown in Table  \ref{tab:contingency}, Table \ref{tab:contingency_sub}, and Fig. \ref{fig:sankey}, but also it can identify the specific cause-and-effect pairs that give the specific causes of the disengagement. The majority of prior studies focusing on the exploration of disengagement causes using taxonomy had successfully concluded the categories \cite{FAVARO2018136}, but they did not identify the specific causes, such as ``a wrong speed control command" or ``software module generated a wrong path and froze" that were valuable to manufacturers in terms of improving the AV system design and testing. 


\section{CONCLUSIONS}
This study created a scalable end-to-end pipeline based on NLP deep transfer learning, which can not only collect, process raw data to generate a consolidated database, but also extract and analyze the causes of AVD from the CA DMV dataset. The best model used in the pipeline was the product of the best practice of transfer learning followed by post-training. A taxonomy covering both human operator trust, AV system safety, and environmental factors was introduced to better understand the causes. Other examples of the results produced from this pipeline were also presented, including AVD counts over the years, word cloud for specific cause categories, statistical analysis between two different types of disengagement initiators. Our model is also able to analyze future incoming AVD in a seamless way to update the current results, which show the potential of our proposed pipeline.

\bibliography{main}



%

\vskip 0pt plus -1fil

\begin{IEEEbiography}[{\includegraphics[width=1in,height=1.55in,clip,keepaspectratio]{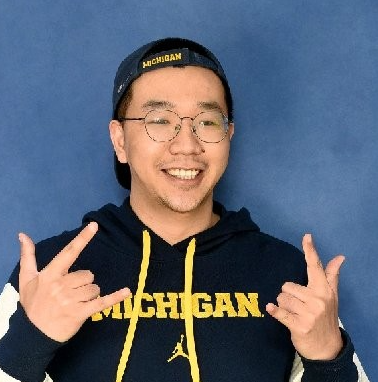}}]{Yangtao Zhang received his B.S. degree in Electrical and Computer Engineering from UM-SJTU Joint Institute, Shanghai Jiao Tong University in 2019 and his master degree in Information Science from University of Michigan, Ann Arbor, in 2021. He is currently a software engineer at Amazon. His main research interests include natural language processing and human-machine interaction.}
\end{IEEEbiography}
\vskip 0pt plus -1fil

\vskip 0pt plus -1fil
\begin{IEEEbiography}[{\includegraphics[width=1in,height=1.55in,clip,keepaspectratio]{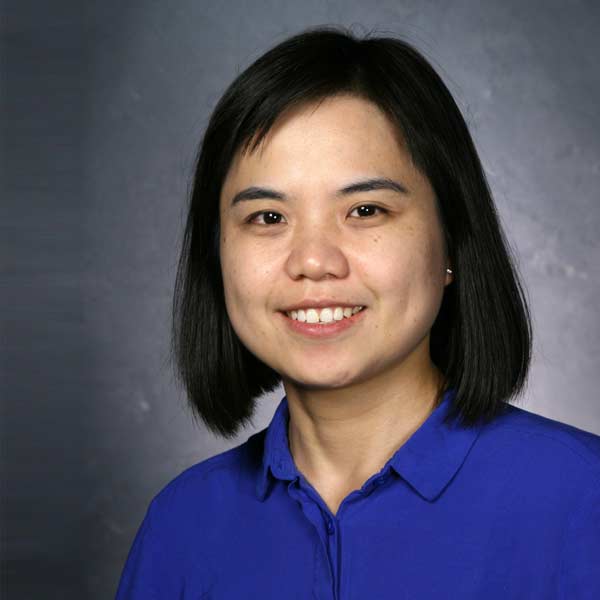}}]{X. Jessie Yang is an Assistant Professor in the Department of Industrial and Operations Engineering, University of Michigan, Ann Arbor. She earned a PhD in Mechanical and Aerospace Engineering (Human Factors) from Nanyang Technological University, Singapore. Dr. Yang’s research include human-autonomy interaction, human factors in high-risk industries and user experience design.}
\end{IEEEbiography}
\vskip 0pt plus -1fil
\begin{IEEEbiography}[{\includegraphics[width=1in,height=1.55in,clip,keepaspectratio]{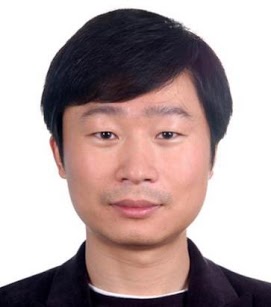}}]{Dr Feng Zhou received the Ph.D. degree in Human Factors Engineering from Nanyang Technological University, Singapore, in 2011 and Ph.D. degree in Mechanical Engineering from Gatech Tech in 2014. He was a Research Scientist at MediaScience, Austin TX, from 2015 to 2017. He is currently an Assistant Professor with the Department of Industrial and Manufacturing Systems Engineering, University of Michigan, Dearborn. His main research interests include human factors, human-computer interaction, engineering design, and human-centered design.}
\end{IEEEbiography}



\end{document}